\documentclass[10pt,twocolumn]{article}
\usepackage[top=2cm, bottom=2cm, left=1.85cm, right=1.85cm]{geometry}
\usepackage{times}  %
\usepackage[sort&compress,numbers]{natbib}  %
\usepackage{graphicx}  %
\usepackage[hyphens]{url}  %
\usepackage{tikzsymbols}
\usepackage{verbatim}
\usepackage{multirow, makecell}
\usepackage{tabularx}
\usepackage{booktabs}
\usepackage{float}
\usepackage{paralist}
\usepackage{varwidth}
\usepackage[small,bf]{caption} %
\usepackage{subcaption} %
\newsavebox\tmpbox

\let\oldbibliography\thebibliography
\renewcommand{\thebibliography}[1]{%
  \oldbibliography{#1}%
  \setlength{\itemsep}{2pt}%
}

\usepackage[hang,flushmargin]{footmisc}

\usepackage[compact]{titlesec}
\titlespacing*{\section}{0pt}{*4}{4pt}
\titlespacing*{\subsection}{0pt}{*3}{3pt}
\usepackage{xspace}

\makeatletter
\def\url@leostyle{%
  \@ifundefined{selectfont}{\def\UrlFont{}}%
  {\def\UrlFont{}}%
}
\makeatother
\usepackage[hyphenbreaks]{breakurl}

\usepackage[bookmarks=true, bookmarksnumbered=true, colorlinks=true, linkcolor=linkcol, citecolor=citecol, urlcolor=urlcol, hypertexnames=true]{hyperref}

\definecolor{darkgreen}{RGB}{0, 100, 0}
\definecolor{linkcol}{rgb}{0.3,0,0}
\definecolor{citecol}{rgb}{0.3,0,0}
\definecolor{urlcol}{rgb}{0.3,0,0}

\newif\ifcomment
\commenttrue

\ifcomment
	\newcommand{\edc}[1]{\textbf{\em\color{red}EDC: #1}}
	\newcommand{\gs}[1]{\textbf{\em\color{blue}GS: #1}}
	
  \newcommand{\sz}[1]{\textbf{\em\color{blue}SZ: #1}}
\else
		\newcommand\edc[1]{}
		\newcommand\gs[1]{}
		{}
    \newcommand\sz[1]{}
\fi
\newcommand{\descr}[1]{\smallskip\noindent\textbf{#1}}

\begin{document}

\title{\bf Feels Bad Man: Dissecting Automated Hateful Meme Detection Through the Lens of Facebook's Challenge}

 \author{
 		Catherine Jennifer\textsuperscript{\rm 1}, Fatemeh Tahmasbi\textsuperscript{\rm 2}, Jeremy Blackburn\textsuperscript{\rm 2},\\[0.5ex] Gianluca Stringhini\textsuperscript{\rm 3}, Savvas Zannettou\textsuperscript{\rm 4}, and Emiliano De Cristofaro\textsuperscript{\rm 1}\\[1.5ex]
\normalsize \textsuperscript{\rm 1}UCL \textsuperscript{\rm 2} Binghamton University \textsuperscript{\rm 3} Boston University \textsuperscript{\rm 4} TU Delft 
 }

\date{}

\maketitle

\begin{abstract}
Internet memes have become a dominant method of communication; at the same time, however, they are also increasingly being used to advocate extremism and foster derogatory beliefs.
Nonetheless, we do not have a firm understanding as to which perceptual aspects of memes cause this phenomenon. 
In this work, we assess the efficacy of current state-of-the-art multimodal machine learning models toward hateful meme detection, and in particular with respect to their generalizability across platforms.
We use two benchmark datasets comprising 12,140 and 10,567 images from 4chan's ``Politically Incorrect'' board (/pol/) and Facebook's Hateful Memes Challenge dataset to train the competition's top-ranking machine learning models for the discovery of the most prominent features that distinguish viral hateful memes from benign ones. 
We conduct three experiments to determine the importance of multimodality on classification performance, the influential capacity of fringe Web communities on mainstream social platforms and vice versa, and the models' learning transferability on 4chan memes.

Our experiments show that memes' image characteristics  provide a greater wealth of information than its textual content. 
We also find that current systems developed for online detection of hate speech in memes necessitate further concentration on its visual elements to improve their interpretation of underlying cultural connotations, implying that multimodal models fail to adequately grasp the intricacies of hate speech in memes and generalize across social media platforms. 
\end{abstract}

\section{Introduction}
Social networking sites have facilitated communication among users worldwide, connecting like-minded individuals who share similar values and enabling the formation of online communities.
As methods of discourse on the Web undergo continuous change to enhance the comprehensibility of personal opinions, so does the potential to advocate hateful beliefs.
Most notably, Internet memes are used as a way of communicating such concepts in an engaging manner, with the most viral of memes acquiring the most attention and becoming near impossible to moderate~\cite{roose_2021}. 
There is a rising interest in developing approaches to better manage this problem; for instance, in 2020, the Hateful Memes Challenge was launched by Facebook AI~\cite{fb_ai_memes_challenge_2020}, aiming to improve the detection of hate speech in multimodal memes.

In spite of the efforts made in multimodal Artificial Intelligence (AI), this problem extends beyond the modalities of a meme; rather, there is very little focus on the visual characteristics which make such content attractive enough to obtain a large quantity of resubmissions on social networks.
The human brain can interpret an image in a mere 13 milliseconds~\cite{trafton_2014}, and image memes have made it convenient for users to quickly comprehend its connotative message. 
Moreover, the circulation of viral hateful memes on the Internet occurs by the influence of different social platforms on each other~\cite{origins_of_memes}. 
However, the Hateful Memes Challenge introduce a dataset with memes that have been generated \emph{artificially}, whereby, such samples do not accurately capture the characteristics of hateful memes that originate and spread on other platforms, limiting the generalizability of these results.

In this paper, we focus on predicting the dissemination of toxic image memes by running experiments on memes from 4chan's Politically Incorrect Board (/pol/) and Facebook's Hateful Memes Challenge using Vision and Language (V\&L) machine learning models to evaluate the potency of multimodal machine learning classification for viral hateful memes. %
Overall, we identify and tackle the following research questions:
\begin{enumerate}
\item How significant is the influence of multimodality in image memes? 
\item How portable are models trained on Facebook's challenge memes on other social platforms?
\item What are the characteristics of hateful viral memes? 
\end{enumerate}

\descr{Methodology.} We start with performing three experiments involving four V\&L classifiers and using Kiela et al.'s challenge dataset for multimodal classification~\cite{kiela_hateful_memes}, and a set of hateful memes collected from /pol/ by Zannettou et al.~\cite{origins_of_memes}. 
More specifically:
\begin{itemize}
\item We use methods by Velioglu and Rose~\cite{velioglu_detectron} to train a VisualBERT model on Zannettou et al.'s dataset to assess the importance of text in hateful meme images. 
\item We focus on the portability of Kiela et al.'s samples on other social networks by evaluating the prediction performance of a UNITER model (with the settings from Muennighoff~\cite{muennighoff_vilio}) on 4chan memes. 
\item We use three models -- UNITER, OSCAR, and an ensemble classifier -- each of which are trained, optimized, and tested only on samples from 4chan to evaluate the generalizability of the Hateful Memes Challenge's best learning algorithms.
\item We conduct a feature analysis to inspect the visual attributes with the most influential impact on the classification accuracy of classifiers from the first and third experiments to discover indicators of virality. 
\end{itemize}

\descr{Findings.} Our main findings can be summarized as follows:

\begin{enumerate}
\item The visual characteristics of memes offer a plethora of information to effectively communicate the image's intended meaning without the inclusion of text. 
This is evident from the model's ability to correctly identify hateful memes 80\% of the time in both unimodal and multimodal representations.
\item The Hateful Memes Challenge dataset is not adequately representative of multimodal hate speech to support the creation of detection algorithms, as demonstrated by the second experiment, when the classifier is evaluated on samples from /pol/.
\item We find four principal characteristics associated with virality in hateful memes: subject matter, facial expressions, gestures, and proportion. In general, hateful viral memes incorporate two or more of these attributes, which is evident from the capacity of the best classifier across all three experiments to correctly classify 84\% of viral memes from 4chan as hateful.
\end{enumerate}

\descr{Remarks.} In this paper, we use the following definition of \emph{hate} for our investigation: ``speech or expression that denigrates a person or persons on the basis of (alleged) membership in a social group identified by attributes such as race, ethnicity, gender, sexual orientation, religion, age, physical, or mental disability, and others,'' as per~\cite{hate_curtis_2016}. 
Also, please be advised that this paper contains uncensored hateful images, which might be disturbing to some readers. 

\begin{table}[t] 
\small
\setlength{\tabcolsep}{2.75pt}
\small
\centering
\begin{tabular}{@{}rrrrrrr} 
 \toprule
&&   & \multicolumn{2}{c}{\textbf{4chan}} &
 \multicolumn{2}{c}{\textbf{Facebook}}\\ 
&& \textbf{\#Memes}
   & \textbf{Hateful} & \textbf{Non-Hateful} & \textbf{Hateful} & \textbf{Non-Hateful} \\
   \midrule
&1A & 8,923  & 3,442 & 0 & 0 & 5,481\\ 
&1B & 8,259 & 2,778 & 0 & 0 & 5,481\\ 
&2 & 10,251  & 750 & 1,001 & 3,019 & 5,481\\
&3 & 2,596  & 1,297 & 1,299 & 0 & 0\\
 \bottomrule
\end{tabular}
\caption{\label{tab:table_datasets}Summary of datasets.}
\label{table:tab1}
\end{table}

\section{Background}
\subsection{Facebook Hateful Memes Challenge}
The Hateful Memes Challenge was launched by Facebook AI to support the development of autonomous systems for the recognition of harmful multimodal content~\cite{fb_ai_memes_challenge_2020}. For this purpose, Kiela et al. proposed a challenge set comprised of multimodal memes conveying hateful or non-hateful messages, constructed in a fashion which makes it difficult for unimodal classifiers to effectively discriminate between the two classes. In particular, samples containing contradictory meanings through their modalities, i.e., “benign confounders” are included in the dataset such that only multimodal models are able to accurately interpret their communicative signals for better predictability.  

\subsection{4chan}
4chan is an anonymous image-sharing board widely recognized for its user's radical opinions and influence on other social media sites.
Particularly members from the /pol/ board that majorly harbor content promoting far-right, misogynistic, and transphobic views which has impacted the information ecosystem and sparked widespread controversy, e.g., the 2016 US presidential elections that flourished the creation of antisemitic memes on /pol/ to advance an agenda of white supremacy~\cite{kek_2017,zannettou2020quantitative}.
In fact, many hateful viral memes trace back to 4chan as its source of origin, with small fringe communities such as /pol/ having the potential to spread such content on larger, more mainstream platforms (e.g., Twitter)~\cite{origins_of_memes}.    

\section{Datasets}

In this section, we introduce the two benchmark datasets used throughout the experimentation, of which is summarized in Table \ref{table:tab1}. We focus on 4chan's /pol/ community and Facebook in this study; thus, we use 10,567 images collected by Ling et al.~\cite{origins_of_memes} as a baseline for the analysis of toxic meme virality. This dataset consists of both multimodal and unimodal samples which are either hateful or non-hateful.  

Additionally, we use the Hateful Memes Challenge dataset created by Kiela et al.\cite{kiela_hateful_memes} considering that Facebook is the most commonly used networking platform to date~\cite{datareportal_2021} and has the potential to exert social influences on the Web ecosystem at mass, making the spread of memes more prominent on the service. Moreover, Facebook's challenge set comprises 12,140 examples of multimodal hate speech that expresses socio-cultural information through is visual modes which makes it suitable for evaluative purposes.
In the rest of the paper, the two datasets are divided into four subsets and used for the three experiments as follows:\smallskip 
\begin{compactenum}
\item A set of 5,481 multimodal non-hateful images from Facebook merged with 3,442 multimodal hateful images from 4chan, and another set of 5,481 multimodal non-hateful Facebook images merged with 2,778 unimodal hateful 4chan images.

\item A set of 1,001 non-hateful and 750 hateful 4chan images with text.

\item A set of 1,299 non-hateful and 1,297 hateful 4chan images with text.\smallskip
\end{compactenum}

\descr{Pre-Processing.}
Facebook's hateful memes competition provided their participants with meme images' extracted text in separate JSON Lines (.jsonl)~\cite{json_lines} formatted files for training, validation, and testing procedures named as 'train.jsonl', 'dev\_unseen.jsonl' and 'test\_unseen.jsonl', respectively. Each image file name in Kiela et al.'s dataset serves as a reference to match the textual and visual modalities of a sample prior to its classification~\cite{driven_data_2020}, leading to the creation of train, validation, and test .jsonl files adapted specifically for each 4chan sample set. We use the Optical Character Recognition (OCR) Python package known as EasyOCR~\cite{easy_ocr} to extract text from 4chan image memes and include the textual content next to the corresponding file's ID in the .jsonl files. 

To ascertain that the extracted text was precisely as depicted in its original meme image, and avoid providing the experimental models with distorted input that would affect its prediction performance, every JSON line in the newly created files was carefully inspected for the manual correction of text arrangement errors or slang words that were not entirely captured by EasyOCR. 

\begin{figure*}[t]
  \centering
  \begin{subfigure}[b]{.3\columnwidth}
    \centering
    \includegraphics[width=\linewidth]{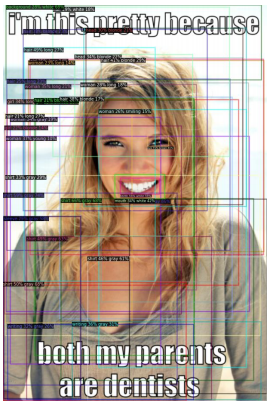}
    \caption{}
    \label{fig:fig1a}
  \end{subfigure}
  \hspace*{0.2cm}
  \begin{subfigure}[b]{1.25\columnwidth}
    \centering
    \includegraphics[width=\linewidth]{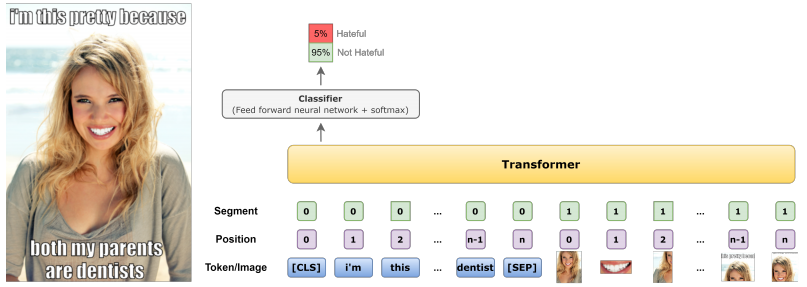}
    \caption{}
    \label{fig:fig1b}
  \end{subfigure}
  \caption{VisualBERT CC ensemble model implementation: (a) an example of Detectron image feature extraction using 36 bounding boxes, (b) an example image meme (left), and a demonstration of the model's architecture and classification process (right) from Velioglu and Rose~\cite{velioglu_detectron}.}
\end{figure*}

\begin{figure*}[t]
\centering
\includegraphics[scale=0.45]{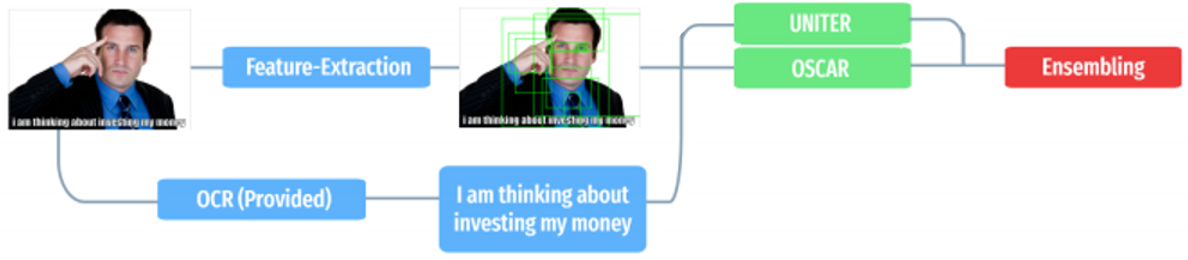}
\caption{Average-prediction ensemble architecture. \copyright Getty Images.}   
\label{fig:fig2}
\end{figure*}

\begin{table*}[t] \normalsize 
\small
\centering
\begin{tabular}{ l|rrrr|rrrr} 
\toprule
 & \multicolumn{4}{c|}{\textbf{Multimodal Memes}} & \multicolumn{4}{c}{\textbf{Unimodal Memes}}\\ \cline{2-9}
 \textbf{Classifier} & \textbf{Precision} & \textbf{Recall} & \textbf{F1} & \textbf{AUC} & \textbf{Precision} & \textbf{Recall} & \textbf{F1} & \textbf{AUC}\\
 \midrule
 VisualBERT CC ensemble & 0.66  & 0.80 & 0.73 & 0.85 & 0.70 & 0.69 & 0.69 & 0.81\\
 \bottomrule
\end{tabular}
\caption{\label{tab:table_sub_exs}Results for The Effects of Multimodality in Hateful Memes.} 
\label{table:tab2}
\end{table*}

\begin{figure*}[t]
  \centering
  \begin{subfigure}[b]{.7\columnwidth}
    \centering
    \includegraphics[width=\linewidth]{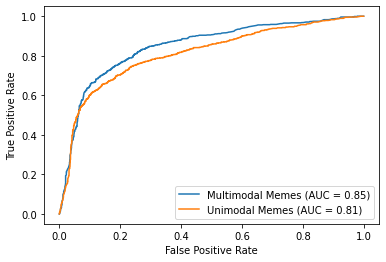}
    \caption{}
    \label{fig:fig3a}
  \end{subfigure}
  \hspace*{0.2cm}
  \begin{subfigure}[b]{.7\columnwidth}
    \centering
    \includegraphics[width=\linewidth]{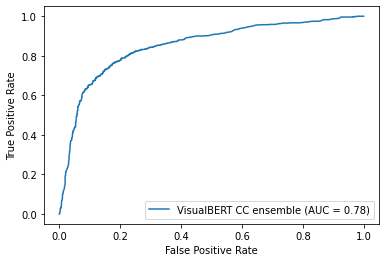}
    \caption{}
    \label{fig:fig3b}
  \end{subfigure}
  \caption{ROC curves for: (a) Multimodal Memes and Unimodal Memes, (b) VisualBERTCC ensemble classifier trained only on Facebook samples.}
\end{figure*}

\begin{table}[t]
\small
\centering
    \begin{tabular}{l|rrrr}
 \toprule
 \textbf{Classifier} & \textbf{Precision} & \textbf{Recall} & \textbf{F1} & \textbf{AUC}\\
 \midrule
 UNITER & 0.54  & 0.24 & 0.33 & 0.56\\
 \bottomrule
\end{tabular}
 \caption{Results for Determining the Generalizability of Facebook's Meme Samples.}
\label{table:tab3}
\end{table}

\begin{figure}[t]
\centering
\includegraphics[scale=0.44]{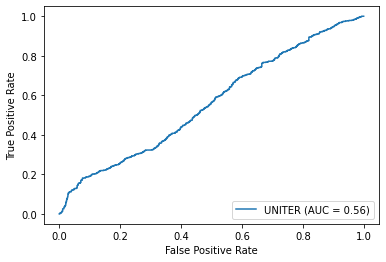}  
\caption{ROC curve for UNITER.} 
    \label{fig:fig4}
\end{figure}

\begin{table}[t] \normalsize 
\small
\centering
\setlength{\tabcolsep}{4pt}
\begin{tabular}{l|rrrrr} %
 \toprule
 \textbf{Classifier} & \textbf{Precision} & \textbf{Recall} & \textbf{F1} & \textbf{AUC}\\
 \midrule
 UNITER & 0.98  & 0.93 & 0.95 & 0.99\\
 OSCAR & 0.95  & 0.95 & 0.95 & 0.99\\
 Average-Prediction Ensemble & 0.96  & 0.96 & 0.96 & 0.99\\
 \bottomrule
\end{tabular}
\caption{\label{tab:tab4}Results for Evaluating Vision-Language Models on Toxic Viral 4chan Memes.}
\label{table:tab4}
\end{table}

\begin{figure*}[t]
  \centering
  \begin{subfigure}[b]{.65\columnwidth}
    \centering
    \includegraphics[width=\linewidth]{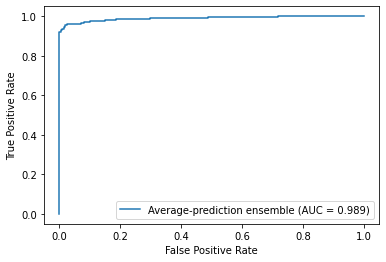}
    \caption{Average-prediction ensemble}
    \label{fig:fig5a}
  \end{subfigure}
  \hspace*{0.2cm}
  \begin{subfigure}[b]{.65\columnwidth}
    \centering
    \includegraphics[width=\linewidth]{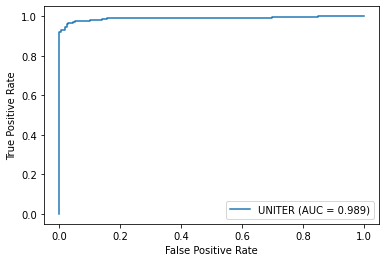}
    \caption{UNITER}
    \label{fig:fig5b}
  \end{subfigure}
    \hspace*{0.2cm}
  \begin{subfigure}[b]{.65\columnwidth}
    \centering
    \includegraphics[width=\linewidth]{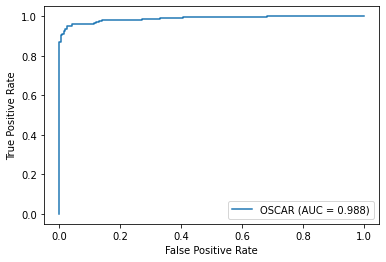}
    \caption{OSCAR}
    \label{fig:fig5c}
  \end{subfigure}
  \caption{ROC curves for third experiment. AUC values for average-prediction ensemble, UNITER, and OSCAR classifiers.}
\end{figure*}

\section{Experimental Setup}
In this section, we describe the procedures undertaken by three classification experiments, and the metrics used to evaluate the performance of each learning model.

\subsection{The Effects of Multimodality in Hateful Memes} 
We consider two cases in the first experiment to obtain a greater insight into the impact of multimodality on model predictability:  the importance of text in image memes, and the effect of unimodality in image memes for accurate classification. The focal point of this experiment is to test whether toxic viral 4chan memes that comprise multiple visual modes have enhanced influential potential on Facebook memes as opposed to those comprising a single communication mode, and allow us to further improve our understanding of how moderately sized fringe Web communities influence mainstream platforms. The first case will henceforth be referred as \emph{Multimodal Memes}, and the second case will be referred as \emph{Unimodal Memes}.

\descr{Train, Validate, and Test Splits.} For \emph{Multimodal Memes}, we benchmark the second-best ranking model of the challenge -- a VisualBERT CC majority-vote ensemble~\cite{velioglu_detectron} -- on the dataset consisting of 8,923 image memes with text, culminating a balanced data distribution of 3,442 hateful 4chan memes and 5,481 non-hateful Facebook memes to form the training set. For better comparative analysis, we maintain a similar class label distribution to the Hateful Memes Challenge dataset of 85\%, 5\%, and 10\% for training, validation, and testing, respectively. Likewise, \emph{Unimodal Memes} uses the dataset comprising 2,778 hateful 4chan image memes without text and 5,481 non-hateful Facebook images memes with text for training and tuning the VisualBERT classifier with close distribution to Kiela et al.'s train set.

\descr{Model implementation.} We perform feature extraction using the object detection algorithm known as \emph{Detectron}~\cite{detectron_2018} to capture important patterns in image memes and enhance the learning ability and generalizability of the classifier. Specifically, we use the \emph{Mask RCNN} deep neural network~\cite{maskrcnn_2018} based on the \emph{ResNet-152} architecture~\cite{he2015deep} to extract features from 100 bounding boxes per image (refer to Figure \ref{fig:fig1a} for an example illustration of a processed sample). We then perform a Hyperparameter Search on multiple VisualBERT CC derivatives to discover a combination of the most optimal parameters for training and select 27 classifiers with the highest ROC-AUC scores on the validation set from~\cite{kiela_hateful_memes}. Finally, a majority-vote approach is taken to combine each model's estimations on Kiela et al.'s test set and form a single ensemble classifier. Figure 1b depicts an overall visual interpretation of this procedure (note that both sub-experiments undergo the same process).

\subsection{Determining the Generalizability of Facebook's Meme Samples} 
In this experiment, we study the influential potential of hateful memes from mainstream social media platforms on comparatively small Web communities. Specifically, we assess a UNITER model \cite{muennighoff_vilio} on the 4chan test set after being trained on Kiela et al.'s dataset. We chose UNITER given its adequacy for Natural Language Processing (NLP) tasks and taking into account that all samples used for this second experiment contain embedded text. 

\descr{Train, Validate, and Test Splits.} All image memes in the 4chan test set for this study are multimodal, with 750 being hateful and 1,001 non-hateful; however, we do not alter the train and validation sets from~\cite{kiela_hateful_memes}. Considering that the majority of memes distributed on social media are benign, a larger portion of samples in the test set are non-hateful to simulate a realistic depiction of how well the model would perform upon deployment on Web platforms and thus end up with an imbalanced class distribution for testing. 

\descr{Model implementation.} Like in the first experiment, we use Detectron to extract image features from memes, but consider 36 bounded boxes per image instead of 100, since UNITER outperforms VisualBERT \cite{muennighoff_vilio} with fewer parameters \cite{chen2020uniter}.
We fine-tune the classifier on the validation set using a binary cross-entropy loss function to compare its probability predictions against true class labels, and use the Adam optimizer~\cite{adam_opt_2014} with the same hyperparameter settings defined in~\cite{adequacy_ma_2021} to train the model for five epochs. 

\subsection{Evaluating Vision-Language Models on Toxic Viral 4chan Memes}
Finally, we evaluate the classification performance of three models, namely UNITER, OSCAR \cite{muennighoff_vilio}, and an average-prediction ensemble formed by computing the weighted mean of the model's combined predictions on multimodal 4chan samples to contrast their capacity to distinguish hateful memes from benign ones.

\descr{Train, Validate, and Test Splits.} We split the dataset for this study into three subsets of which two will be used for training and optimization, and another for testing. To account for the moderate quantity of samples used in this experiment, we follow a 70:10:20 data split such that precisely 1,997, 199, and 400 image memes with text are used for training, validation, and testing, respectively. Unlike the first and second experiments, this dataset excludes samples containing long text due to the constraint of maximum 512 tokens imposed by transformer models~\cite{bert_2019}. 

\descr{Model implementation.} To introduce diversity in the models' predictions and reduce feature redundancy, we extract feature vectors from various Regions of Interests (RoIs) in image memes by defining different quantities of bounding boxes. 
Then, the UNITER and OSCAR models undergo the same procedure for training and optimization as UNITER in \emph{The Effects of Multimodality in Hateful Memes} experiment.  A third, ensemble classifier is created from the two aforementioned by averaging their individual predictions upon completion of their training (refer to Figure \ref{fig:fig2} for an illustration of the average-prediction model's classification of an example image).

\subsection{Metrics}
The four following measures are used to assess the performance of the models: Precision, Recall, F1-Score, and the Area Under the Curve of the Receiver Operating Characteristic (AUC-ROC).  Precision and Recall are widely recognized as effective approaches to determine classification performance on imbalanced datasets, with the F1-Score providing a balance between the two measures to dictate an overall outcome of each classifier's estimation quality on unseen data. We also chose the AUC-ROC metric to compare how well the classifiers are able to discriminate between the classes under different test scenarios.

\section{Results}

\subsection{The Effects of Multimodality in Hateful Memes} 
Performance results for this experiment are shown in Table \ref{table:tab2}. We see that the VisualBERT CC classifier has a higher recall than precision after being tested under the conditions of \emph{Multimodal Memes} based on the fact that the training dataset comprised solely of multimodal samples.  Moreover, this type of model has been developed specifically for V\&L tasks, which has enabled it to generate more truthful estimations on the test set upon learning associations between memes' text and image features. Unsurprisingly, the precision score is greater in \emph{Unimodal Memes} since all samples in Kiela et al.'s dataset contains embedded text causing the classifier to identify a larger quantity of benign memes than hateful ones. A 69\% recall rate nevertheless suggests the predictive capacity of VisualBERT CC is above that of a mediocre one. 

In Figure \ref{fig:fig3a}, we plot the ROC curves for both sub-experiments. An AUC score of approximately 0.80 across the experiments means the VisualBERT classifier can correctly discriminate between hateful and non-hateful memes 80\% of the time, regardless of the modalities involved. \emph{Unimodal Memes} notably reveals that memes do not necessitate text to inflict extremist ideology and increase its potential for online dissemination demonstrating that image characteristics of memes are just as meaningful as those incorporating text. We also plot the ROC curve of the model's classification performance when trained only on Kiela et al.'s training set in Figure \ref{fig:fig3b}, which projects a very close AUC result of 0.78 to that of the sub-experiments, further supporting our findings.

\subsection{Determining the Generalizability of Facebook's Meme Samples} 
We report the results of UNITER's performance in Table \ref{table:tab3}, which shows poor classification performance given its near-chance AUC score (0.56). The ROC curve in Figure \ref{fig:fig4} provides a better interpretation of this outcome. Furthermore, the model has a recall rate of 0.24 on the 4chan test set after being trained on Facebook's dataset, indicating low discriminatory ability between the classes. This implies that Kiela et al.'s dataset may not adequately simulate memes shared on social media considering one of the best performing models cannot generalize well to memes from other social platforms. 

Figure \ref{fig:fig4} further shows that UNITER incorrectly labels many hateful memes as non-hateful, suggesting that it is unable to capture the visual features in test samples at a rate sufficient enough to make truthful classifications, resulting in an inadequate true positive rate (TPR) for addressing the challenge of hate speech recognition in multimodal memes.

\descr{Evaluating Vision-Language Models on Toxic Viral 4chan Memes.} Table \ref{table:tab4} shows the results of each classifier, i.e., UNITER, OSCAR, and average-prediction, attain AUCs of 0.989, 0.988, and 0.989, respectively. We observe that the average-prediction ensemble achieves the greatest recall rate compared with the other two. However, UNITER obtains a greater precision (0.979) than the ensemble model meaning it can correctly label memes as hateful approximately 98\% of the time. Nevertheless, a higher recall is favorable for this classification task and although the average-prediction classifier has the same AUC (0.989) as UNITER, it delivers the best overall performance. We also see that OSCAR is 0.02\% more likely to accurately identify hateful memes than UNITER given its recall of 0.950, but is nonetheless the weakest performing model as this score is impeded by its inferior AUC.

Figures \ref{fig:fig5a}, \ref{fig:fig5b}, and \ref{fig:fig5c} demonstrate ROC curve plots corresponding to UNITER, OSCAR, and the average-prediction model, respectively. Ultimately, UNITER and the average-prediction ensemble have exceptional discriminative ability between both classes with identical potential (considering their AUCs); however, we can discern in Figure \ref{fig:fig5a} that the ensemble classifier achieves a higher TPR making it preferable for this problem.

\section{Related Work}
\descr{Meme propagation.} Previous work has focused on measuring and tracing meme dissemination on the Web. Zannettou et al. \cite{origins_of_memes} introduced a custom metric to measure the visual similarity between image memes to track variants of meme families from polarized Web communities such as 4chan's /pol/, Gab, and The\_donald, in an effort to study their impact on meme propagation and analyze the influential correlations between the social networking platforms.    

\descr{Indicators of viral image memes.} Arturo Deza and Devi Parikh \cite{understanding_image_virality} conducted a semantic evaluation of the perceptual cues in viral memes, identifying 5 key attributes that link to virality: 'Animal', 'Synthgen', 'Beautiful', 'Explicit', and 'Sexual' - Each of which elicit different emotional reactions from its viewers and potentially affect their decision to share a post.  

\descr{Detection of hateful and offensive memes.} Kiela et al. \cite{kiela_hateful_memes} introduced a challenge dataset of 10,000 artificially generated multimodal memes representative of real ones publicized on social platforms, and annotated as hateful or non-hateful. Various approaches to this competition have been tried, including the use of early fusion strategy with transformer models to combine the visual elements and textual content of memes prior to their classification \cite{zhu_2020_enhance, muennighoff_vilio, velioglu_detectron, lippe_kingsterdam, vlad_sand}.

\descr{\emph{Novelty:}} The detection of hateful speech in multimodal memes \cite{kiela_hateful_memes} is the most similar work to ours thus far. However, this work is the first to consider the anticipation of such content prior to its publication - viral hateful memes in particular, which become extremely challenging to moderate once posted. We also identify limitations in Kiela et al.'s dataset and the approaches used by winning contestants of the Hateful Memes Challenge.  

\section{Conclusion}

\subsection{Limitations}

As previously mentioned, hateful viral memes from fringe Web communities such as 4chan's /pol/ also appear in comic strip format (e.g., Tyrone~\cite{kym_tyrone}). The training set from our \emph{Evaluating Model Generalization on Toxic Viral 4chan Memes} experiment contains 119 of 1,997 images which comprise of multiple panels depicting a popular meme subject, with almost half of the TPs produced by the average-prediction classifier showing this attribute. 

Although seemingly minor, the study of memes composed of more than a single panel is worthy of consideration to try to comprehend how memes such as the Tyrone comic series was successful in gaining mass resubmissions and imitations on social media.
Unfortunately, we could not examine this particular element closely enough to deem it as another possible indicator of virality given none of Kiela et al.'s~\cite{kiela_hateful_memes} data samples are viral due to their nature of construction. 
Thus, further investigation in the context of spatial vicinity is necessary to establish whether the presence of viral meme subjects enhance the virality potential of multiple-panel image memes. 

Another aspect of this study is the examination of how many panels in a comic meme strip would be too many, provided each panel displays text to illustrate a story, and an abundance of text reduces a meme's online influence (as shown from our experimental outcomes).
However, there is currently limited availability of meme datasets for such studies and thus we hope that future work in this area will contribute to their development.

Moreover, careful creation and scrutiny of train, validation, and test datasets is very time consuming, %
consequently limiting the number of available samples for experimentation and excluding images depicting long multipaned story illustrations reducing sampling diversity. 

We also encountered GPU compatibility issues due to the fact that learning algorithms from the Hateful Memes Challenge necessitate appropriate CUDA versions to operate. 
Hence, we could not use the first ranked classifier for our study given its high CUDA version requirement for our experimentation 
environment. Needless to say, it is worthy to continue this exploratory research using methods by~\cite{zhu_2020_enhance} to observe their model's learnability. 

Finally, it is important to note that our sociocultural identity has a strong influence on our understanding of online content. 
For instance, the ethnic background of one individual may cause a hateful perception toward an image meme, but perhaps not by another individual. 
This differentiation in points of view has arguably been the greatest challenge in our work thus far and is evident in the experimental classifier's biased tendency to label samples containing certain terms (e.g., Jew) as hateful when in reality such terms are also used in non-hateful contexts. 
Nevertheless, we hope that future work will study the visual characteristics of meme images to better interpret the true intentions of their creators. 

\subsection{Main Take-Aways}

This paper presented a multimodal deep learning approach to determine whether advancements made toward the detection of hateful memes by the Hateful Memes Challenge and the solutions thereof generalize to 4chan and other fringe communities. 
Our experiments showed that the inclusion of text in image memes does not significantly impede the spread of extremist views, given the very close classification scores obtained by the models when evaluated on unimodal memes. 

We found that Kiela et al.'s challenge dataset~\cite{kiela_hateful_memes} does not realistically depict actual memes shared on social media, which has resulted in the development of learning algorithms that are incapable of adequately recognising hateful memes from other social networking platforms. 

Our results also attested to the effectiveness of ensemble V\&L classifiers for enhancing detection performance. 

Overall, our work provides a first step toward assessing the viability of state-of-the-art multimodal machine learning methods, in an effort to improve the creation and deployment of autonomous systems for hate speech detection in memes posted on the Web.

\descr{Acknowledgements.} We gratefully acknowledge the support of NetSySci Research Laboratory, Cyprus University of Technology for allowing us to use their NVIDIA server drivers for our experiments.

\bibliographystyle{abbrv}
\bibliography{bibfile.bib}%

\begin{figure}[t]
  \centering
  \begin{subfigure}[b]{.25\columnwidth}
    \centering
    \includegraphics[width=\linewidth]{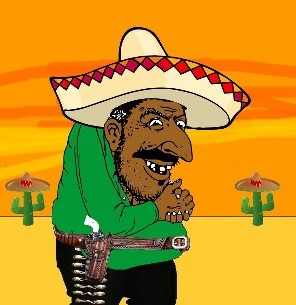}
    \caption{}
    \label{fig:fig6a}
  \end{subfigure}%
  \hspace*{0.2cm}
  \begin{subfigure}[b]{.25\columnwidth}
    \centering
    \includegraphics[width=\linewidth]{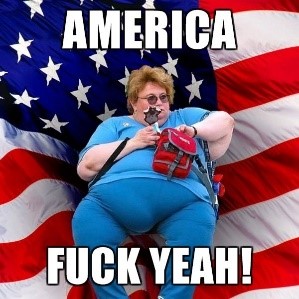}
    \caption{}
    \label{fig:fig6b}
  \end{subfigure}%
  \\
    \begin{subfigure}[b]{.23\columnwidth}
    \centering
    \includegraphics[width=\linewidth]{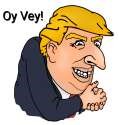}
    \caption{}
    \label{fig:fig6c}
  \end{subfigure}%
  \hspace*{0.2cm}
  \begin{subfigure}[b]{.18\columnwidth}
    \centering
    \includegraphics[width=\linewidth]{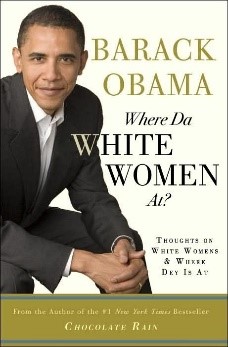}
    \caption{}
    \label{fig:fig6d}
  \end{subfigure}%
  \caption{National stereotypes and famous figures: (a) Mexican, (b) American, (c) Donald Trump caricature, and (d) Barack Obama from Zannettou et. al.'s dataset~\cite{origins_of_memes}.}
  \label{fig:fig6}
\end{figure}

\begin{figure}[t]
\small
  \centering
  \begin{subfigure}{.17\columnwidth}
    \centering
    \includegraphics[width=\linewidth]{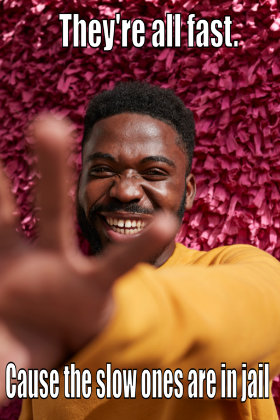}
    \caption{}
    \label{fig:fig7a}
  \end{subfigure}%
  \hspace*{0.2cm}
  \begin{subfigure}{.19\columnwidth}
    \centering
    \includegraphics[width=\linewidth]{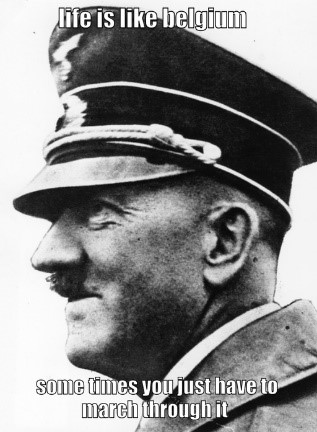}
    \caption{}
    \label{fig:fig7b}
  \end{subfigure}%
  \hspace*{0.2cm}
    \begin{subfigure}{.17\columnwidth}
    \centering
    \includegraphics[width=\linewidth]{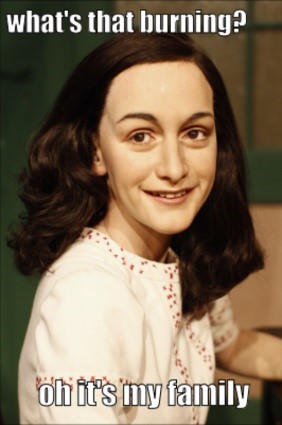}
    \caption{}
    \label{fig:fig7c}
  \end{subfigure}%
  \caption{Subject of hateful memes (a) an African American stereotype, (b) Adolf Hitler, and (c) Anne Frank from the HM dataset~\cite{kiela_hateful_memes}. \copyright Getty Images}.
  \label{fig:fig7}
\end{figure}

\begin{figure}[t]
  \centering
  \begin{subfigure}[b]{.4\columnwidth}
    \centering
    \includegraphics[width=\linewidth]{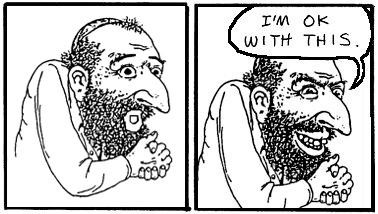}
    \caption{}
    \label{fig:fig8a}
  \end{subfigure}
  \hspace*{0.2cm}
  \begin{subfigure}[b]{.2\columnwidth}
    \centering
    \includegraphics[width=\linewidth]{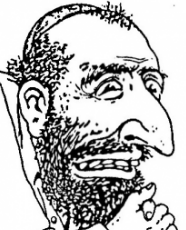}
    \caption{}
    \label{fig:fig8b}
  \end{subfigure}
\\
  \begin{subfigure}[b]{.35\columnwidth}
    \centering
    \includegraphics[width=\linewidth]{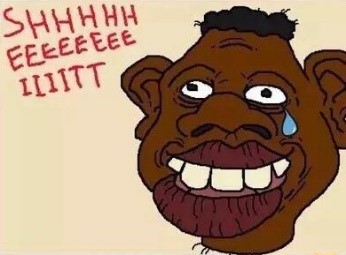}
    \caption{}
    \label{fig:fig8c}
  \end{subfigure}
      \hspace*{0.2cm}
  \begin{subfigure}[b]{.28\columnwidth}
    \centering
    \includegraphics[width=\linewidth]{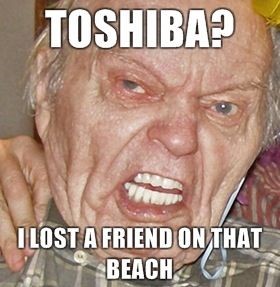}
    \caption{}
    \label{fig:fig8d}
  \end{subfigure}
  \caption{Emotions portrayed through facial expressions: (a) surprise and malevolence, (b) fear, (c) sadness, and (d) anger from Zannettou et. al.'s dataset~\cite{origins_of_memes}.}
  \label{fig:fig8}
\end{figure}

\begin{figure}[t]
  \centering
  \begin{subfigure}[b]{.21\columnwidth}
    \centering
    \includegraphics[width=\linewidth]{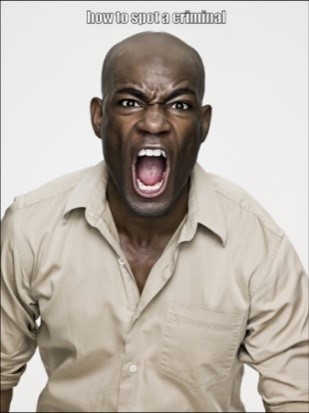}
    \caption{}
    \label{fig:fig9a}
  \end{subfigure}
  \hspace*{0.2cm}
  \begin{subfigure}[b]{.21\columnwidth}
    \centering
    \includegraphics[width=\linewidth]{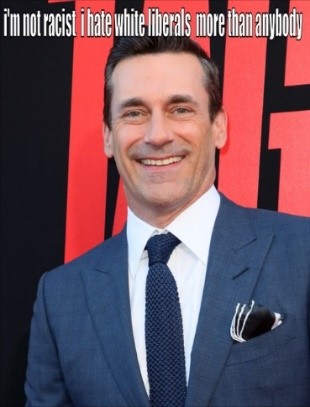}
    \caption{}
    \label{fig:fig9b}
  \end{subfigure}
\\
  \begin{subfigure}[b]{.37\columnwidth}
    \centering
    \includegraphics[width=\linewidth]{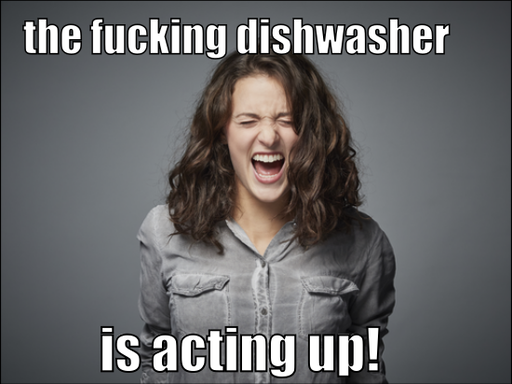}
    \caption{}
    \label{fig:fig9c}
  \end{subfigure}
  \caption{TP test samples portraying: (a) enragement, (b) happiness, and (c) frustration from the HM dataset~\cite{kiela_hateful_memes}. \copyright Getty Images.} 
  \label{fig:fig9}
\end{figure}

\begin{figure}[t]
  \centering
  \begin{subfigure}{.43\columnwidth}
    \centering
    \includegraphics[width=\linewidth]{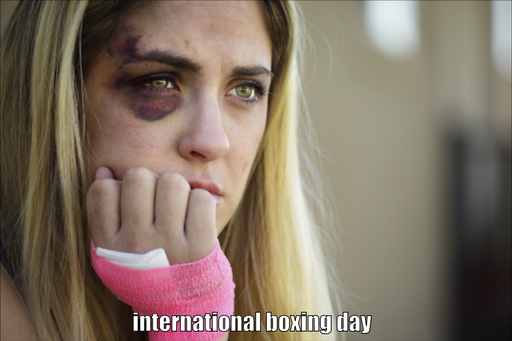}
    \label{fig:fig10a}
  \end{subfigure}
  \hspace*{0.4cm}
  \begin{subfigure}{.4\columnwidth}
    \centering
    \includegraphics[width=\linewidth]{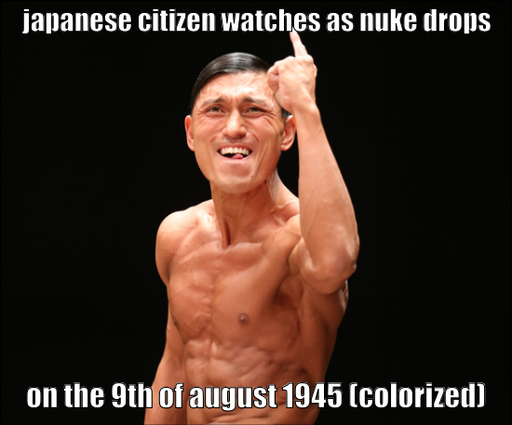}
    \label{fig:fig10b}
  \end{subfigure}
  \caption{Misclassified samples from Multimodal Memes, from the HM dataset~\cite{kiela_hateful_memes}. \copyright Getty Images.}
  \label{fig:fig10}
\end{figure}

\begin{figure}[t]
  \centering
  \begin{subfigure}{.26\columnwidth}
    \centering
    \includegraphics[width=\linewidth]{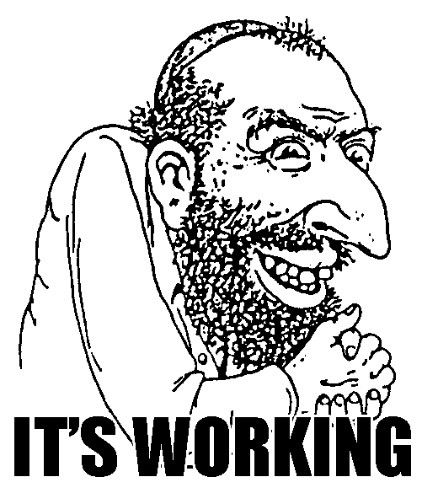}
    \caption{}
    \label{fig:fig11a}
  \end{subfigure}
  \hspace*{0.4cm}
  \begin{subfigure}{.42\columnwidth}
    \centering
    \includegraphics[width=\linewidth]{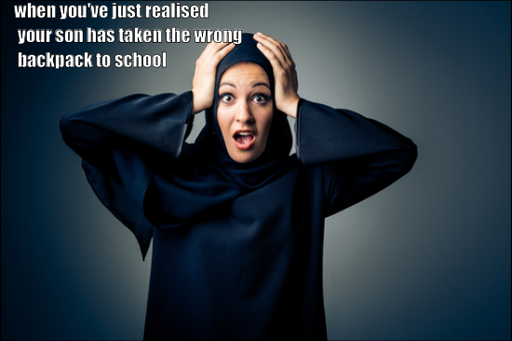}
    \caption{}
    \label{fig:fig11b}
  \end{subfigure}
  \caption{Example of: (a) textual meaning influenced by gestural behavior, (b) correctly predicted sample containing contradictory modalities, from Zannettou et. al's dataset and the HM dataset~\cite{origins_of_memes, kiela_hateful_memes}. \copyright Getty Images.}
  \label{fig:fig11}
\end{figure}

\begin{figure}[t]
  \centering
\begin{subfigure}{.33\columnwidth}
    \centering
    \includegraphics[width=\linewidth]{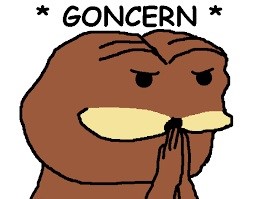}
  \end{subfigure}
  \begin{subfigure}{.23\columnwidth}
    \centering
    \includegraphics[width=\linewidth]{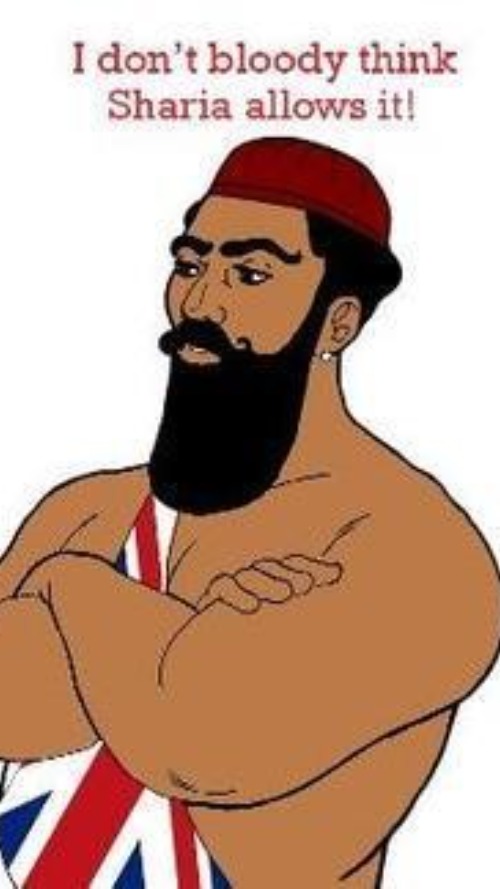}
  \end{subfigure}
    \begin{subfigure}{.28\columnwidth}
    \centering
    \includegraphics[width=\linewidth]{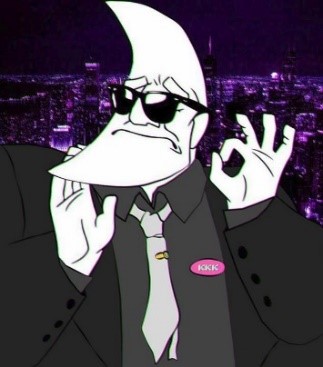}
  \end{subfigure}
    \caption{Example close-up viral hateful meme images from Zannettou et. al.s dataset~\cite{origins_of_memes}.  \copyright Getty Images.}
    \label{fig:fig12}
\end{figure}

\begin{figure}[t]
    \begin{subfigure}{.23\columnwidth}
    \centering
    \includegraphics[width=\linewidth]{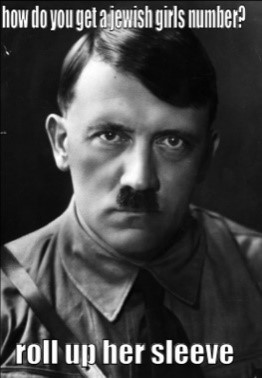}
  \end{subfigure}
  \hspace*{0.4cm}
    \begin{subfigure}{.23\columnwidth}
    \centering
    \includegraphics[width=\linewidth]{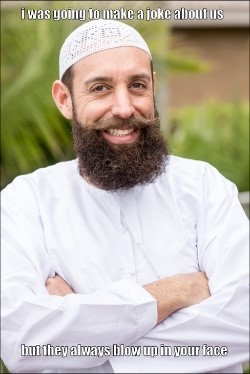}
  \end{subfigure}
  \hspace*{0.4cm}
    \begin{subfigure}{.33\columnwidth}
    \centering
    \includegraphics[width=\linewidth]{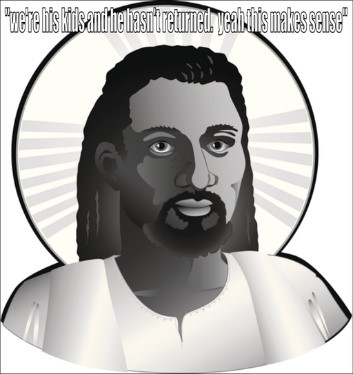}
  \end{subfigure}
  \caption{TPs made by VisualBERT ensemble classifier, from the HM dataset~\cite{kiela_hateful_memes}. \copyright Getty Images.}
  \label{fig:fig13}
\end{figure}

\appendix
\section{Feature Importance \& Virality}
In this section, we conduct a formal analysis of the characteristics of hateful memes that contribute to its virality potential with respect to the most prominent features recognized by the two best performing classifiers from the first and third experiments (the VisualBERT CC and average prediction ensemble models). Hereby, we discuss the top four features learned by the models leveraging Know Your Meme -- the largest encyclopedia of Internet memes -- as a guideline for the affirmation of memes' virality.    

\begin{enumerate}
\item \textbf{Subject matter:} 67\% of viral memes in the training set and 55\% of true positive (TP) classifications made by the average-prediction ensemble depict a character, stereotype representation, caricature, or famous individual. This suggests that images containing a region of primary focus (or emphasis) have a greater likelihood of becoming viral. We find this to be the case as the subject of a hateful meme image is indicative of its target audience (see example Figure \ref{fig:fig6}). Likewise, the VisualBERT CC classifier correctly predicts 599 and 520 hateful memes from sub-experiment test sets 1 and 2, respectively. Although Kiela et al.'s dataset does not consist of viral memes, we still see that the model's performance was influenced by racial stereotype portrayals (e.g., Figure \ref{fig:fig7a}) and impactful historical figures (as shown in Figures \ref{fig:fig7b} and \ref{fig:fig7c}), further supporting this finding.

\item \textbf{Facial expressions:} Image subjects that portray emotions through their facial expressions strongly impacted classification decisions made by the VisualBERT CC classifier (93.7\% of viral hateful memes in the training set, and 84\% of TPs displayed this feature), showing that meme virality is influenced by the expression of sentiment to advocate beliefs. The majority of viral hateful 4chan memes used in the first experiment depict subjects which convey emotions through facial expressions (e.g., Figure \ref{fig:fig8}), and 87\% of hateful test samples were classified correctly by the VisualBERT CC model while demonstrating this attribute (see Figure \ref{fig:fig9}). Peculiarly, the two hateful memes shown in Figure \ref{fig:fig10} were falsely predicted as non-hateful when tested under the condition of \emph{Unimodal Memes}, implying that image features can be as informative as text for the anticipation of a viral meme.

\item \textbf{Gestures:} Gestures of meme subjects are on par with their facial expressions with regard to feature importance. This attribute also indicates underlying connotations of an image to change the entirety of its meaning (e.g., the text in Figure 11a alone is not hateful yet demonstrates antisemitic undertones when interpreted with the stereotypical illustration of a Jewish man malevolently rubbing his hands together). Subsequently, subjects' gestural behavior is considered by the average-prediction model before producing a final prediction solely based on its textual meaning (93.7\% and 84\% of viral hateful memes in the train and test sets possess this trait together with the top 1 feature). Similarly, the VisualBERT CC classifier uses this characteristic to assess each test sample in its entirety prior to classification (refer to Figure \ref{fig:fig11b} for an example of a correctly labelled sample indicating this attribute). 

\item \textbf{Proportion:} The majority of hateful viral memes possessing two or more of the above-mentioned features use a close-up shot such as those shown in Figure \ref{fig:fig12} (70\% and 84\% of viral hateful memes in the training sets used for the first and third experiments tightly frame their subjects of focus – the same is the case for 29\% and 55\% of TPs made by the VisualBERT CC and average-predictions models). We argue that meme authors depict the full form of the image's figure to convey their message more clearly through facial emotional expressions and gestures. Results from \emph{The Effects of Multimodality in Hateful Memes} show that the VisualBERT CC model picks up on this feature to distinguish between the two classes for Kiela et al.'s test samples (see Figure \ref{fig:fig13} for TP prediction examples). 

\end{enumerate}

\end{document}